\documentclass{aa}

\pdfoutput=1 
\usepackage{txfonts}
\usepackage{graphicx}
\usepackage{color}
\usepackage{natbib}
\usepackage{rotating}
\usepackage{url}
\bibpunct{(}{)}{;}{a}{}{,} 

\begin{document}

\title{Modelling the spinning dust emission from dense interstellar clouds}
\author{N. Ysard\inst{1}
\and M. Juvela\inst{1}
\and L. Verstraete\inst{2}}
\offprints{Nathalie Ysard, \email{nathalie.ysard@helsinki.fi}}
\institute{Department of Physics, PO Box 64, 00014 University of Helsinki, Finland
\and Institut d'Astrophysique Spatiale (IAS), Universite Paris-Sud, 91405 Orsay, France}
\abstract
{Electric dipole emission arising from rapidly rotating Polycyclic Aromatic Hydrocarbons (PAHs) is often invoked to explain the anomalous microwave emission. This assignation is based on i) an observed tight correlation between the mid-IR emission of PAHs and the anomalous microwave emission; and ii) a good agreement between models of spinning dust and the broadband anomalous microwave emission spectrum. So far often detected at large scale in the diffuse interstellar medium, the anomalous microwave emission has recently been studied in detail in well-known dense molecular clouds with the help of Planck data.}
{While much attention has been given to the physics of spinning dust emission, the impact of varying local physical conditions has not yet been considered in detail. Our aim is to study the emerging spinning dust emission from interstellar clouds with realistic physical conditions and radiative transfer.}
{We use the DustEM code to describe the extinction and IR emission of all dust populations. The spinning dust emission is obtained with SpDust, which we have coupled to DustEM. We carry out full radiative transfer simulations and carefully estimate the local gas state as a function of position within interstellar clouds.}
{We show that the spinning dust emission is sensitive to the abundances of the major ions ($\ion{H}{ii}$, $\ion{C}{ii}$) and we propose a simple scheme to estimate these abundances. We also investigate the effect of changing the cosmic-ray rate. In dense media, where radiative transfer is mandatory to estimate the temperature of the grains, we show that the relationship between the spinning and mid-IR emissivities of PAHs is no longer linear and that the spinning dust emission may actually be strong at the centre of clouds where the mid-IR PAH emission is weak. These results provide new ways to trace grain growth from diffuse to dense medium and will be useful for the analysis of anomalous microwave emission at the scale of interstellar clouds.}
{}
\authorrunning{N. Ysard}
\titlerunning{Modelling the spinning dust emission from dense interstellar clouds}
\maketitle

\section{Introduction}
\label{introduction}

Discovered in the nineties, the anomalous microwave emission (AME) has aroused great interest \citep{Kogut1996, Leitch1997}. First because it appears in a frequency window that is optimal for the detection of the Cosmic Microwave Background (CMB) fluctuations. \citet{DL98} proposed that AME could be caused by electric dipole emission of rapidly rotating grains: the spinning dust emission. This mechanism is now most often invoked to explain the AME and several models have been published \citep{Ali2009, Ysard2010a, Hoang2010, Silsbee2011}. The study of spinning dust could help in understanding the life cycle of interstellar dust grains because it may be a new tracer of the smallest grains, the interstellar Polycyclic Aromatic Hydrocarbons (PAHs).

The preference of spinning dust models over other mechanisms is based on several arguments. First, AME is correlated with dust IR emission and this correlation is particularly tight for the mid-IR emission of small grains. Second, AME is weakly polarized as expected for PAHs because these grains are not supposed to be aligned with the interstellar magnetic field \citep{Battistelli2006, Casassus2008, Lopez2011}. Third, the shape and the intensity of the AME can be reproduced with spinning dust spectra (e.g. Watson et al. 2005; Planck Collaboration et al. 2011b, and many other references). However, the spinning dust emission depends on the local physical conditions (gas ionisation state and radiation field) and on the size distribution of small grains. Recent observations of interstellar clouds point out dissimilar morphologies in the mid-IR and in the microwave range that may be explained by local variations of the environmental conditions \citep{Casassus2006, Casassus2008, Ysard2010b, Castellanos2011, Vidal2011}. In this work we study the spinning dust emission of interstellar clouds including a treatment of the gas state and radiative transfer. In this context we reexamine the relationship between the AME and the dust IR emission.

The paper is organised as follows. In Section \ref{models} we describe the models. In Section \ref{gas_properties} we detail our method to estimate the gas properties (ionisation state and temperature). In Section \ref{environment} we present the variations of the spinning dust spectrum with the gas density and with the intensity of the radiation field. We also consider variations of the cosmic-ray ionisation rate as suggested by recent observations. In Section \ref{radiative_transfer} we present the spinning dust emission with radiative transfer modelling. Finally, we present in Section \ref{conclusions} our conlusions.

\section{Models}
\label{models}

Current models of spinning dust \citep{DL98, Ali2009, Ysard2010a, Hoang2010, Silsbee2011} take into account a number of processes for the rotational excitation and damping of the grains: the emission of IR photons, the collisions with neutral and ionised gas particles (\ion{H}{i}, \ion{H}{ii}, and \ion{C}{ii}), the plasma drag (\ion{H}{ii} and \ion{C}{ii}), the photoelectric effect, and the formation of H$_2$ molecules at the surface of the grains. The publically available SpDust\footnote{Available at \url{http://www.tapir.caltech.edu/∼yacine/spdust/spdust.html}} code \citep{Ali2009, Silsbee2011} includes the most recent developments regarding the gas-grain interactions and the grain dynamics (rotation around non-principal axis of inertia). The results of SpDust agree well with other models that include a more detailed treatment of the IR emission or of the gas-grain interactions \citep{Ysard2010a, Hoang2010}.  SpDust is fast and well-suited for coupling to other codes, especially radiative transfer codes. In the following, we use SpDust to model the spinning dust emission.

In order to estimate dust emission from the mid-IR to the microwave range in a consistent way, we coupled SpDust with the dust emission model described by \citet{Compiegne2011}, DustEM\footnote{Available at \url{http://www.ias.u-psud.fr/DUSTEM}.}. DustEM is based on the formalism of \citet{Desert1990} and includes three dust types: interstellar PAHs, amourphous carbonaceous grains, and amourphous silicates. We used the dust populations defined by \citet{Compiegne2011} for the diffuse, high galactic latitude interstellar medium (DHGL). For PAHs we assumed a log-normal size distribution with centroid $a_0=0.64$ nm and width $\sigma=0.4$, with a dust-to-gas mass ratio $M_{PAH}/M_H=7.8\times 10^{-4}$.

In current models, the smallest grains (PAHs) carry the spinning dust emission that is sensitive to the gas density and the radiation field intensity\footnote{$G_0$ is a scaling factor for the radiation field integrated between 6 and 13.6 eV. The standard radiation field corresponds to $G_0 = 1$ and to an intensity of $1.6\times 10^{−3}$ erg/s/cm$^{−2}$ \citep{Mathis1983}.}, $G_0$, but also to the ionisation state (abundance of the $\ion{H}{ii}$ and $\ion{C}{ii}$ ions, noted $x_H$ and $x_C$ respectively). Radiative transfer calculations are performed with the CRT\footnote{Available at \url{http://wiki.helsinki.fi/display/~mjuvela@helsinki.fi/CRT}.} tool \citep{Juvela2003, Juvela2005}, to which we have coupled DustEM and SpDust. CRT is only used to estimate the dust temperature and the resulting dust emission from the mid-IR to the microwave range. Our treatment of the gas properties is presented in Section \ref{gas_properties}.

\section{Gas state}
\label{gas_properties}

As discussed above, the dynamics of spinning dust grains involves gas-grain interactions and radiative processes. The spinning dust emission is therefore sensitive to the gas density ($n_H$) and temperature ($T_{{\rm gas}}$), and to the intensity of the UV radiation field traced by the factor $G_0$. In particular the gas-grain interactions depend on the gas ionisation state, i.e., the abundance of the major charged species (electrons, \ion{H}{ii}, \ion{C}{ii}, ect.), which primarily depends on $n_H$, $T_{{\rm gas}}$, and $G_0$ but also on the chemistry occurring locally. Realistic modelling consequently requires a consistent treatment of the spinning motion of the grains and gas ionisation state. For the present work, where we consider the influence of radiative transfer on the spinning dust emission (see Section \ref{radiative_transfer}), we treat the gas ionisation state with a simplified scheme that we present below. Using this scheme, we then discuss the influence of $n_H$ and $G_0$, and look at the effect of an enhanced cosmic-ray ionisation rate as suggested by recent observations (see Section \ref{environment}).

When the radiation field intensity is low ($G_0 \leqslant 1$), inelastic collisions with neutral and ionised species of the interstellar gas become the dominant processes for the excitation and the damping of the grain rotation \citep{Ali2009, Ysard2010a}. The ion fractions $x_H = n_{\ion{H}{ii}}/n_H$ and $x_C = n_{\ion{C}{ii}}/n_H$ where $n_H = n(\ion{H}{i}) + n(\ion{H}{ii}) + 2n({\rm H}_2)$ accordingly need to be carefully determined to perform a quantitative study of the variations of spinning dust emission with environmental properties. Where CO has not formed (unshielded regions in which most of the gas phase carbon is in the form of \ion{C}{ii} or \ion{C}{i}), we estimate the electron and ion fractions ($x_e=n_e/n_H$, $x_H$, and $x_C$) by simultaneously solving the \ion{H}{i} / \ion{H}{ii} and \ion{C}{i} / \ion{C}{ii} equilibria, including the recombination of carbon with H$_2$ \citep{Roellig2006}. Furthermore, we take into account the recombination of \ion{C}{ii} with PAHs as described in \citet{Wolfire2008}.

In neutral gas and neglecting the contribution of helium, the ionisation balance of hydrogen including H$_2$ reads
\begin{eqnarray}
 ({\rm \ion{H}{i}, H_2}) + {\rm CR} & \rightleftarrows & ({\rm \ion{H}{ii}, H_2^+}) + {\rm e}^{\,-} \\
 \zeta_{CR} (1-x_H) & = & x_H x_e n_H a_H,
\end{eqnarray}
where $\zeta_{CR}$ is the cosmic-ray ionisation rate per second and per proton and $a_H = 3.5 \times 10^{-12} (T/300 {\rm K})^{-0.75}$ cm$^3$/s is the \ion{H}{ii} recombination rate \citep{Roellig2006}. Unless otherwise stated, we assume $\zeta_{CR} = 5 \times 10^{-17}$ s$^{-1}$H$^{-1}$. In regions where CO has not formed, we assume that \ion{C}{ii} is the dominant ionised heavy element and write the electron fraction as $x_e \simeq x_H + x_C$. The \ion{C}{ii} abundance thus becomes
\begin{equation}
\label{premiere_estimation}
 x_C = x_e - \frac{1}{1+x_e n_H a_H / \zeta_{CR}}.
\end{equation}
On the other hand, $x_C$ can be derived from the ionisation balance of carbon where we take into account the following reactions:
\begin{eqnarray}
 {\rm \ion{C}{i}} + h\nu & \stackrel{k_i}{\longrightarrow} & {\rm \ion{C}{ii}} + {\rm e}^{\,-} \\
 {\rm \ion{C}{ii}} + {\rm e}^{\,-} & \stackrel{k_r}{\longrightarrow} & {\rm \ion{C}{i}} \\
 {\rm \ion{C}{ii}} + {\rm PAH}^-/{\rm PAH} & \stackrel{k_x}{\longrightarrow} & {\rm \ion{C}{i}} + {\rm PAH}/{\rm PAH}^+ \\
 {\rm \ion{C}{ii}} + {\rm H}_2 & \stackrel{k_a}{\longrightarrow} & {\rm CH_2}^+ + h\nu \;\; {\rm or} \;\; {\rm CH}^+ + {\rm \ion{H}{i}}.
\end{eqnarray}
The rate coefficients $k_i$, $k_r$ and $k_x$ are taken from \citet{Wolfire2008}. From the database of the Meudon PDR code \citep{LePetit2006}, we take
\begin{equation} 
k_a\simeq 1.7\,10^{-15}+1.5\,10^{-10}\exp(-4640/T_{{\rm gas}})$ cm$^3$ s$^{-1},
\end{equation}
where the two terms are for the CH$_2^+$ and CH$^+$ products, respectively. The C/\ion{C}{ii} ionisation balance then reads
\begin{equation}
 ([{\rm C}] - x_C) k_i = x_C x_e n_H k_r + x_C x_{{\rm PAH}} n_H k_x + x_C y n_H k_a / 2,
\end{equation}
where $y = 2n_{H_2}/n_H$ is the H$_2$ fraction and $[{\rm C}] = n_C/n_H = n(\ion{C}{i})/n_H + x_C$ is the total carbon abundance, which we take to be $1.3\times 10^{-4}$. This leads to an expression for $x_C$:
\begin{equation}
\label{seconde_estimation}
 x_C = [{\rm C}] \left(  1 + x_e \frac{n_H k_r}{k_i} + x_{{\rm PAH}} \frac{n_H k_x}{k_i} + y \frac{n_H k_a}{2 k_i} \right)^{-1}.
\end{equation}
Combining Eqs. \ref{premiere_estimation} and \ref{seconde_estimation} yields a third-degree equation for $x_e$ whose solution is $x_s$. In shielded regions, $x_C$ drops rapidly to form C and then CO. The electron fraction $x_e$ follows this evolution down to a value $x_{dc}$ corresponding to the case of dark clouds where the ionisation is mostly due to cosmic rays. In this case, we assume that the electron fraction follows the formula that \citet{Williams1998} derived from an analysis of C$^{18}$O, H$^{13}$CO$^+$ and DCO$^+$ observations in low-mass cores:
\begin{equation}
x_{dc}= 2000 \sqrt{ \frac{2\zeta_{CR}}{y n_H}},
\end{equation}
where $y=2n_{H_2}/n_H$. We then write $x_e={\rm MAX}(x_{dc}, x_s)$.

\begin{figure}[!t]
\centerline{
\includegraphics[width=0.4\textwidth]{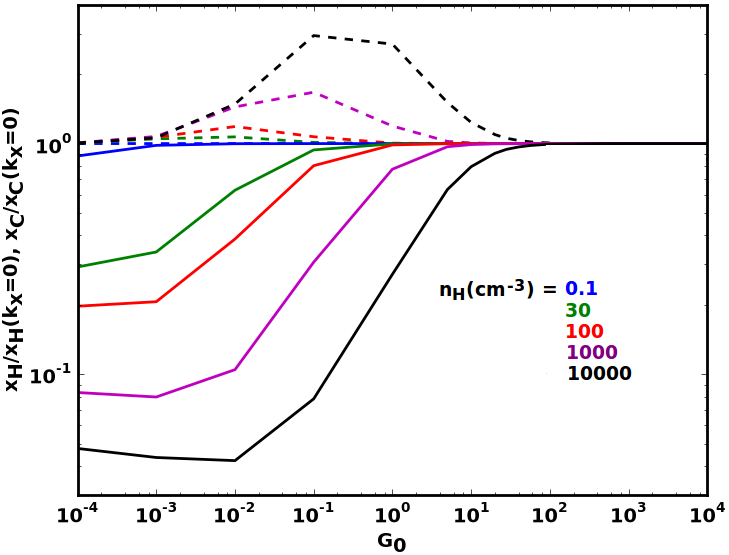}}
\caption{Ratio of ion fractions calculated as decribed in Section \ref{gas_properties} and of ion fractions calculated omitting the recombination of \ion{C}{ii} with PAHs ($k_x = 0$). Results are shown for $n_H = 0.1$ (blue), 30 (green), 100 (red), $1\,000$ (magenta), and $10\,000$ cm$^{-3}$ (black), for $x_H$ (dashed lines from bottom to top) and $x_C$ (solid lines from top to bottom).}
\label{ratio_xH_xC} 
\end{figure}

The \ion{C}{ii} and \ion{H}{ii} fractions are then derived from Eqs. \ref{premiere_estimation} and \ref{seconde_estimation}. Fig. \ref{ratio_xH_xC} shows the influence of the recombination of \ion{C}{ii} with PAHs on the calculation of $x_H$ and $x_C$. When it is omitted ($k_x = 0$), the impact is the strongest when the radiation field intensity is the weakest. The \ion{C}{ii} fraction is over estimated by a factor of up to 20 in dense clouds, whereas the \ion{H}{ii} fraction is underestimated. The recombination with PAHs has no influence when $G_0 \gtrsim 50$, whatever the density.

\begin{figure}[!t]
\centerline{
\includegraphics[width=0.4\textwidth]{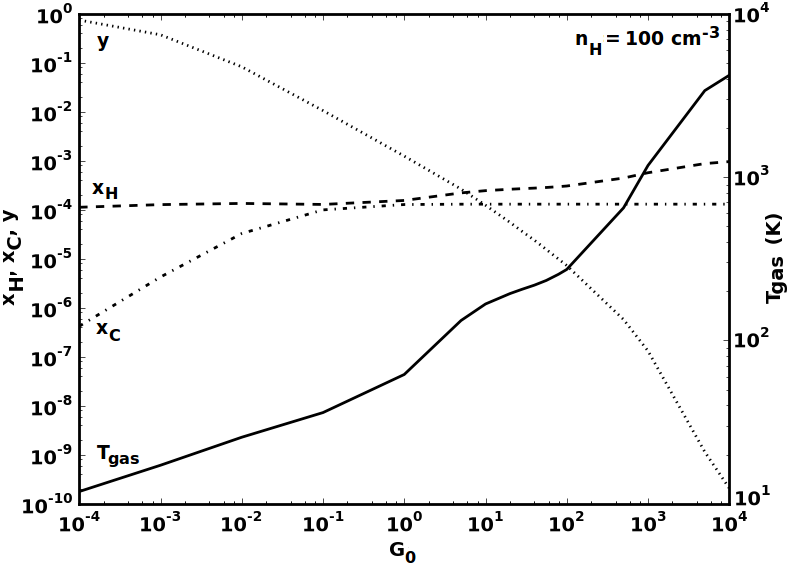}}
\caption{Ion fractions $x_H$ (dashed line) and $x_C$ (dot-dashed line), H$_2$ fraction $y$ (dotted line), and gas temperature $T_{{\rm gas}}$ (full line) as a function of the intensity of the radiation field $G_0$ for $n_H = 100$ H/cm$^{-3}$.}
\label{figure_gas} 
\end{figure}

The above fractions depend on the gas temperature, $T_{gas}$, as well as on the hydrogen molecular fraction $y$. To derive these quantities, we performed a grid of simulations with CLOUDY \citep{Ferland1998}. We then interpolated on these grids to obtain $T_{{\rm gas}}$ and $y$ in the optically thin limit ($N_H\leq 10^{20}$ H/cm$^2$). We note that even if the gas is fully molecular ($y=1$), the reactions of \ion{C}{ii} with H$_2$ have little influence because their rates are much lower than those of reactions involving PAHs. The results are shown in Fig. \ref{figure_gas} for a gas density of $n_H = 100$ cm$^{-3}$.

\section{Influence of local physical conditions}
\label{environment}

\subsection{Variations with the gas density}
\label{gas_density}

\begin{figure}[!t]
\centerline{
\includegraphics[width=0.4\textwidth]{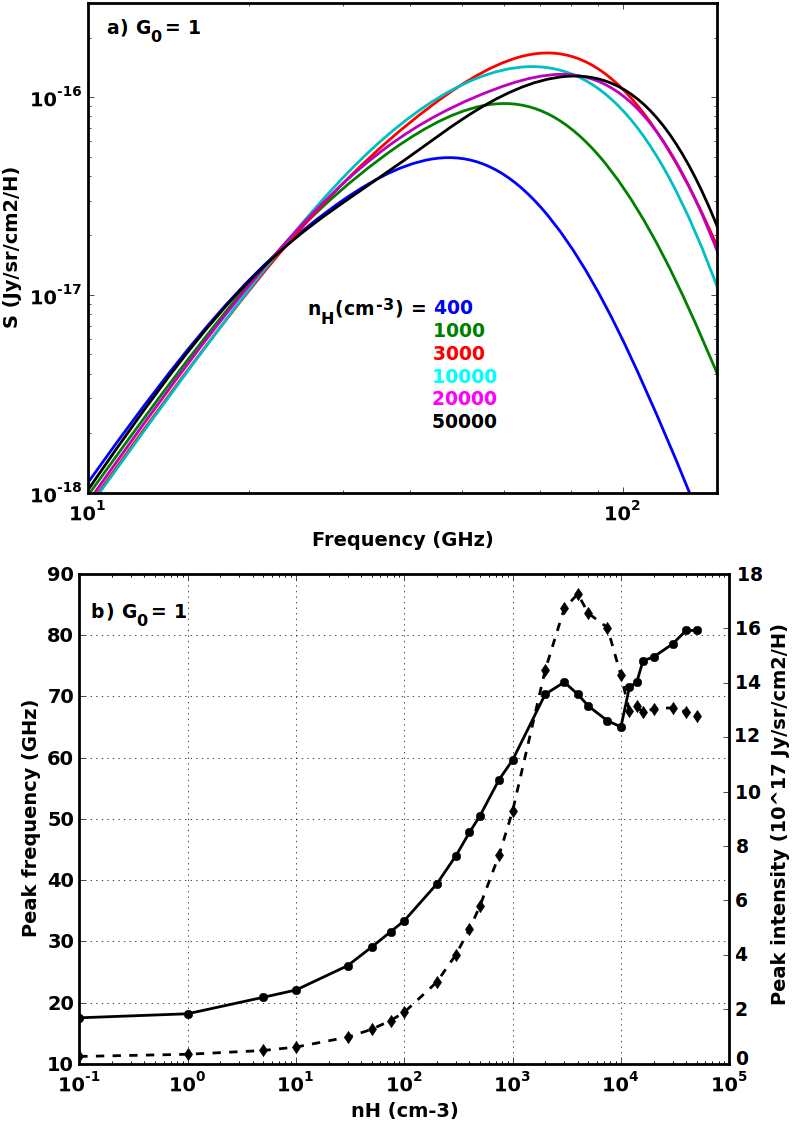}}
\caption{{\it a)} Spinning dust spectra per H column density for interstellar PAHs illuminated by the standard ISRF, $G_0 = 1$, for $n_H = 400, 1000, 3\,000, 4\,000, 7\,500, 20\,000$ and $50\,000$ cm$^{-3}$. {\it b)} Peak frequencies (full line + circles) and intensities per H column density at the peak (dashed line + diamonds) of the spinning dust spectra for $G_0 = 1$ and $n_H = 0.1 - 10^5$ cm$^{-3}$.}
\label{variations_nH} 
\end{figure}

We study the variations of the spinning dust spectrum for grains illuminated by the standard interstellar radiation field, ISRF, with $G_0 = 1$, and for clouds with densities ranging from $n_H = 0.1$ to $10^5$ cm$^{-3}$. For each density, the gas properties are estimated according to $n_H$ and the radiation field intensity, as decribed in Section \ref{gas_properties}. The resulting spectra are shown in Fig. \ref{variations_nH} and we also display the variation of the peak frequency and of the maximum intensity of the spectra as functions of $n_H$.

\begin{figure}[!t]
\centerline{
\includegraphics[width=0.4\textwidth]{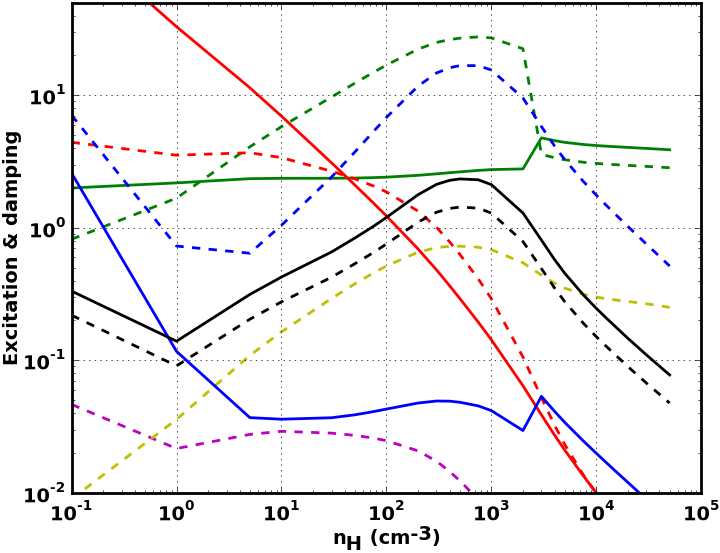}}
\caption{Contribution of the different processes to the damping (solid line) and the excitation (dashed line) of the spinning dust emission for a grain with a radius of 3.5 $\mathring{A}$ illuminated by the standard ISRF with $G_0 = 1$. The contributions are normalized to the drag produced by non-elastic collisions in a pure \ion{H}{i} gas of a density $n_H$ \citep{DL98}. The red lines show the contribution of IR emission, the green lines of collisions with neutrals, the blue lines of collisions with ions, the black lines of plasma drag, the magenta line of photoelectric effect, and the yellow line of H$_2$ formation.}
\label{contributions} 
\end{figure}

When $n_H \lesssim 10$ cm$^{-3}$, the rotational excitation and damping are dominated by the IR emission (Fig. \ref{contributions}). Thus the peak intensity of the spinning dust spectrum does not much depend on the gas properties and it varies little when increasing $n_H$. The slight increase of the peak frequency and of the intensity of the spectra for $n_H \geqslant 1$ cm$^{-3}$ is caused by the increase of the influence of collisions with neutral species (Fig. \ref{contributions}).
If $10 \lesssim n_H \lesssim 4\,000$ cm$^{-3}$, the gas-grain interactions become the dominant processes for rotational excitation and damping. This results in an increased peak frequency when $n_H$ increases. 
For $4\,000 \lesssim n_H \lesssim 10\,000$ cm$^{-3}$, the gas-grain interactions are still the dominant processes but the \ion{H}{ii} and \ion{C}{ii} fractions dramatically decrease when $n_H$ increases. As a result, the ion-grain collisions are less efficient and the peak frequency decreases (Fig. \ref{contributions}).
Finally if $n_H \gtrsim 10\,000$ cm$^{-3}$, the peak frequency increases, while the emissivity varies only little.

The numbers given here are valid only for the ISRF with $G_0 = 1$. If the intensity of the radiation field is increased (decreased), the threshold values from one domain to the next are increased (decreased). We emphasize the importance of correctly estimating the ion fractions when modelling spinning dust emission.

\subsection{Variations with the intensity of the interstellar radiation field}
\label{intensity}

\begin{figure}[!t]
\centerline{
\includegraphics[width=0.4\textwidth]{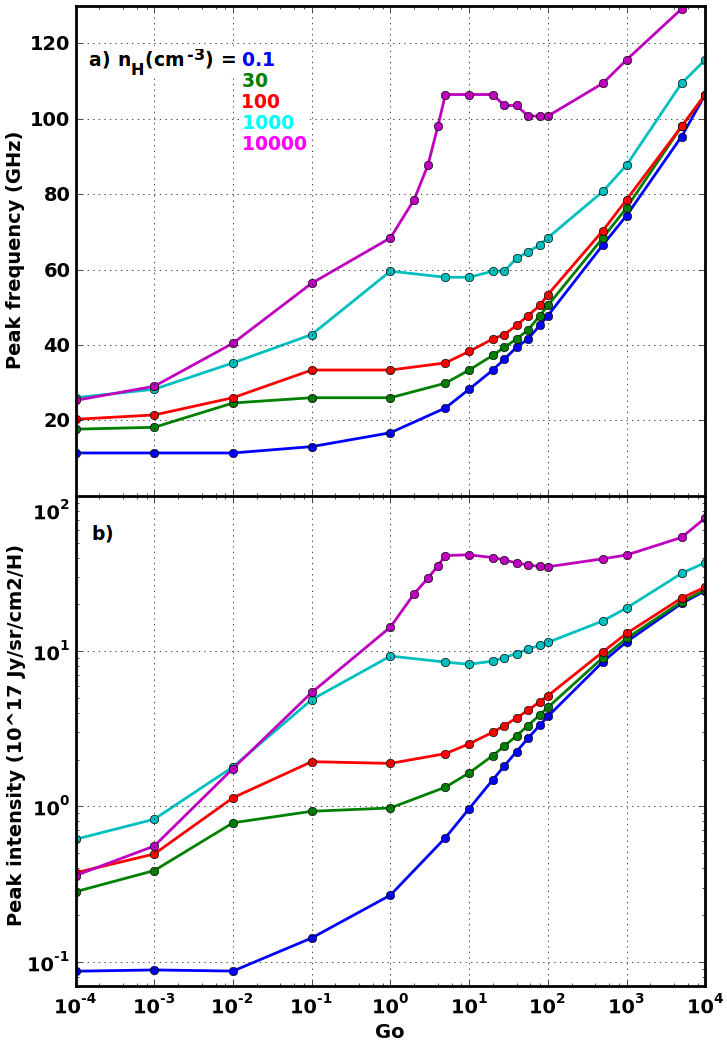}}
\caption{{\it a)} Peak frequencies of the spinning dust spectra for $G_0 = 10^{-4}$ to $10^4$ and $n_H = 0.1$ (blue), 30 (green), 100 (red), $1\,000$ (turquoise), and $10\,000$ cm$^{-3}$ (purple), from bottom to top. {\it b)} Intensities per H column density of the spinning dust spectra at peak frequencies for the same environmental parameters.}
\label{variations_Go} 
\end{figure}

We study the variations of the spinning dust spectrum for grains illuminated by the ISRF scaled by $G_0$ factors between $10^{-4}$ and $10^4$. Several gas densities are considered, from 0.1 to $10^4$ H/cm$^3$. The ion fractions are calculated according to the values of $n_H$ and $G_0$ (see Section \ref{gas_properties}). The results can be seen in Fig. \ref{variations_Go}.

First, for low $G_0$ the dominant processes are gas-grain interactions: the excitation is led by collisions with ions, while the damping is dominated by plasma drag and collisions with neutrals. This explains why the spinning dust spectrum varies very little with $G_0$. Depending on the gas density, this applies to $G_0 \lesssim 10^{-3}-10$. A correct estimate of ion fractions is thus important. For instance, if the recombination of \ion{C}{ii} with PAHs is omitted when calculating $x_H$ and $x_C$ (Fig. \ref{ratio_xH_xC}), the spinning dust spectrum is shifted to higher frequencies for low-$G_0$. The peak shift caused by reaction 6 is between 1 and 25 GHz for $n_H = 30$ to $10\,000$ cm$^{-3}$, while the intensity is increased by 5 to 45\%.
A second domain can be distinguished for $1 \lesssim G_0 \leqslant 100$. The grains become positively charged when $G_0$ increases. For these intermediate radiation field intensities the collisions with ions are still the dominant exciting process even if less efficient. As a result, we see a slight decrease of both the emissivity and the peak frequency of the spinning dust emission, which is more intense when the density is higher.
For $G_0 \geqslant 100$, IR emission dominates the rotational excitation and damping, so that increasing $G_0$ results in increased emissivity and peak frequency.

We stress that the variations of the spinning dust spectrum with $G_0$ are not monotonous and that they depend on the gas density. Moreover, the PAHs mid-IR emission is proportional to $G_0$, so that we do not expect a tight correlation of the IR bands with the AME if the latter is caused by spinning dust emission.

\subsection{Influence of the cosmic-ray ionisation rate $\zeta_{CR}$}
\label{cosmic_rays}

\begin{figure}[!t]
\centerline{
\includegraphics[width=0.4\textwidth]{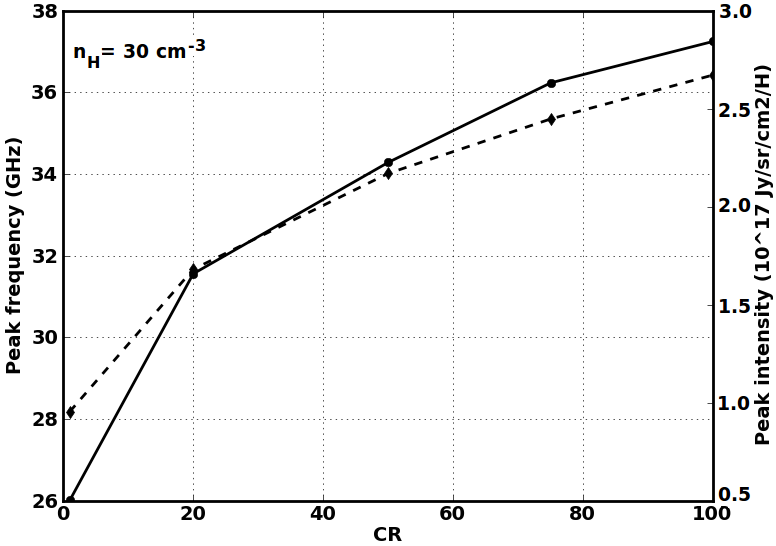}}
\caption{Peak frequencies (solid line + circles) and intensities per H column density at the peak frequencies (dashed line + diamond) of the spinning dust spectra as a function of the cosmic-ray ionisation rate equal to the standard rate $\zeta_{CR} = 5 \times 10^{-17}$ s$^{-1}$ multiplied by the factor CR = [1-100].}
\label{variations_CR} 
\end{figure}

Gas-grain interactions through collisions or plasma drag are one of the dominant processes for the rotational damping and excitation. A proper evaluation of the ion fractions in the gas, $x_H$ and $x_C$, is therefore important for the estimation of the spinning dust emission. The carbon is ionised by the UV radiation field and thus $x_C$ depends on $G_0$ (see Section \ref{gas_properties}). However, because of the Lyman cut of the radiation field, inside clouds the hydrogen is mainly ionised by cosmic-rays and $x_H$ depends on the cosmic-ray ionisation rate $\zeta$. In the previous sections, we adopted the Galactic standard value,  $\zeta_{CR} = 5 \times 10^{-17}$ s$^{-1}$, measured towards dense molecular clouds. However, several authors have inferred an enhanced cosmic-ray ionisation rate from observations of diffuse clouds ($n_H < 500$ cm$^{-3}$). They showed that this rate can exceed 50 times the standard cosmic-ray ionisation rate \citep{McCall2003, Liszt2003, Shaw2006}.

To investigate the influence of the cosmic-ray ionisation rate, we modelled spinning dust spectra for $n_H = 30$ cm$^{-3}$, $G_0 = 1$, and $\zeta = CR \times \zeta_{CR}$ with $1 \leqslant CR \leqslant 100$ (Fig. \ref{variations_CR}). When $\zeta_{CR}$ is increased by two orders of magnitude, $x_H$ is multiplied by 10 and the spinning dust spectrum is shifted by 10 GHz and its intensity is multiplied by three. The cosmic-ray ionisation rate therefore has a significant impact on the spinning dust emission. Its variations as a function of the gas density may produce a more complex behaviour than those presented in Figs. \ref{variations_nH} and \ref{variations_Go}.

\section{Spinning dust emission with radiative transfer}
\label{radiative_transfer}

An interesting result obtained with the Planck data is that there is AME associated with all the interstellar phases of our Galaxy. It is detected in diffuse and dense media, ionised or neutral, and its spectrum can be fitted with a basic spinning dust model in all cases \citep{PlanckMarshall2011, PlanckDickinson2011}. However, at finer angular resolution ($\sim$ a few arcminutes), the morphology of molecular clouds differs in the mid-IR and in the microwave \citep{Castellanos2011, Vidal2011}. We showed above that the spinning dust spectrum is very sensitive to the environmental conditions and we claim that it could explain the observed differences. Indeed, the variations of the gas properties and $G_0$ in dense molecular clouds are strong from the cloud edge to the centre. Hence, dense clouds are excellent targets to test the hypothesis of the spinning dust emission at the origin of the observed AME. Therefore, we investigate the spinning dust emission emerging from dense interstellar clouds using the combination of DustEM and CRT, in addition to SpDust. 

We model starless spherical clouds with density distributions that are suitable to fit interstellar clouds \citep{Arzoumanian2011, Dapp2009}: 
\begin{equation}
n(r) = \frac{n_0}{1 + (r/H_0)^2} \;\;\; {\rm for} \; r \leqslant R_{out},
\end{equation}
with $R_{out}$ the outer radius, $H_0 = R/3$ the flat internal radius, and $n_0$ the central density. The clouds are illuminated by the ISRF extinguished by a visual extinction $A_V =1$. We consider that small and large grains are present everywhere in the clouds with constant abundances. The gas ionisation state is estimated as described in Section \ref{gas_properties}.

\subsection{Gas temperature}

\begin{figure}[!t]
\centerline{
\includegraphics[width=0.4\textwidth]{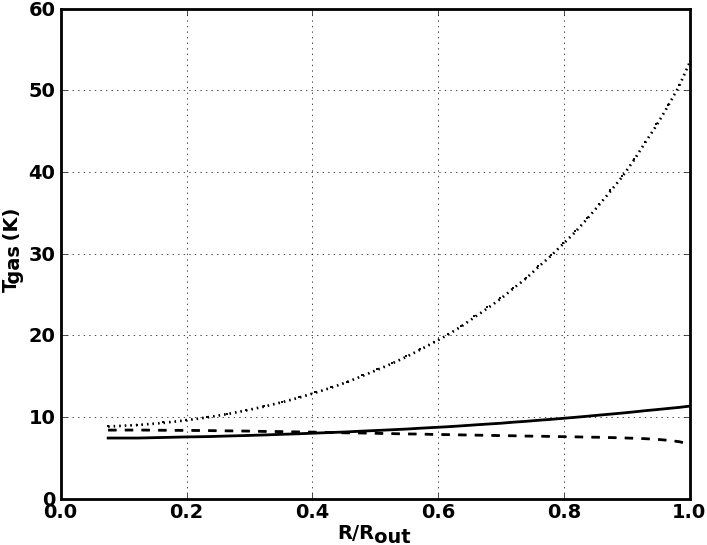}}
\caption{Gas temperature as a function of the cloud radius for clouds with $R_{out} = 0.1$ pc and $n_0 = 10^4$ (solid line) and $10^5$ H/cm$^3$ (dashed line), and for a cloud with $R_{out} = 1$ pc and $n_0 = 10^3$ H/cm$^3$ (dotted line).}
\label{Tgas_profiles} 
\end{figure}

In the dense molecular clouds the gas temperature grid calculated with CLOUDY in the optically thin limit is no longer relevant (see Section \ref{gas_properties}). We used our own radiative transfer code to estimate $T_{{\rm gas}}$ in spherical model clouds. This model takes into account cosmic-ray heating, line cooling, coupling between gas and dust, and photoelectric heating. The calculations follow the description of \citet{Goldsmith2001} with the exception that the line cooling rates are estimated with Monte-Carlo radiative transfer modelling instead of using the large velocity gradient approximation \citep{Juvela2011}. The line transfer is calculated with the abundances listed in \citet{Goldsmith2001} and assuming a turbulent linewidth of 1 km/s. The resulting temperature profiles are shown in Fig. \ref{Tgas_profiles}.

\subsection{Starless molecular cloud}
\label{starless_molecular_cloud}

\begin{figure}[!t]
\centerline{
\includegraphics[width=0.4\textwidth]{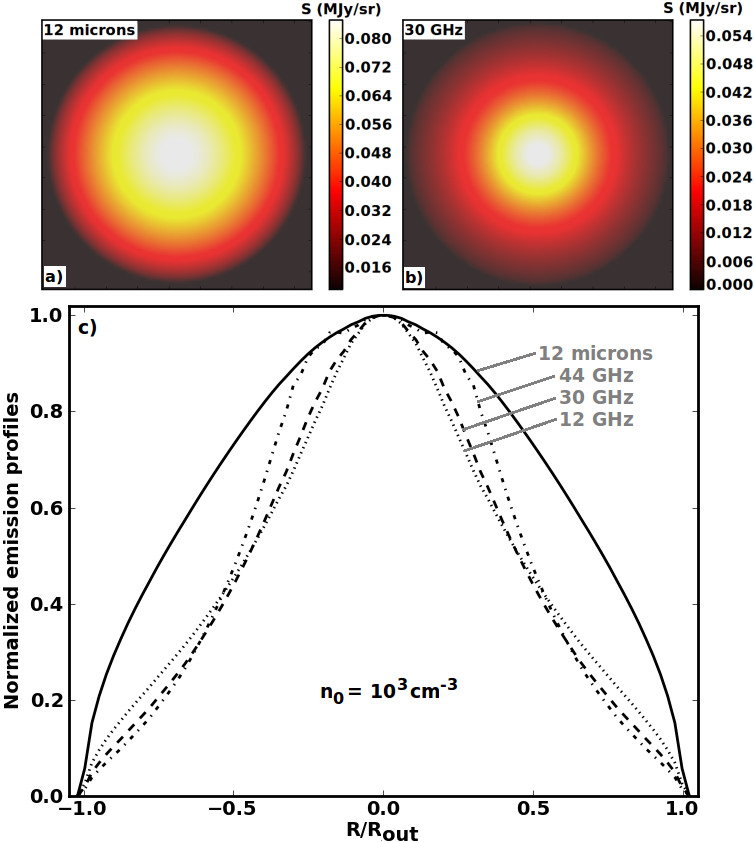}}
\caption{Surface brightness maps at 12 $\mu$m ($a$) and 30 GHz ($b$) and the emission profiles ($c$) at 12 $\mu$m (solid line), 12 GHz (dotted line), 30 GHz (dashed line), and 44 GHz (dot-dashed line) for the starless molecular cloud with $n_0 = 10^3$ cm$^{-3}$.}
\label{figure_1e3} 
\end{figure}

\begin{figure}[!t]
\centerline{
\includegraphics[width=0.4\textwidth]{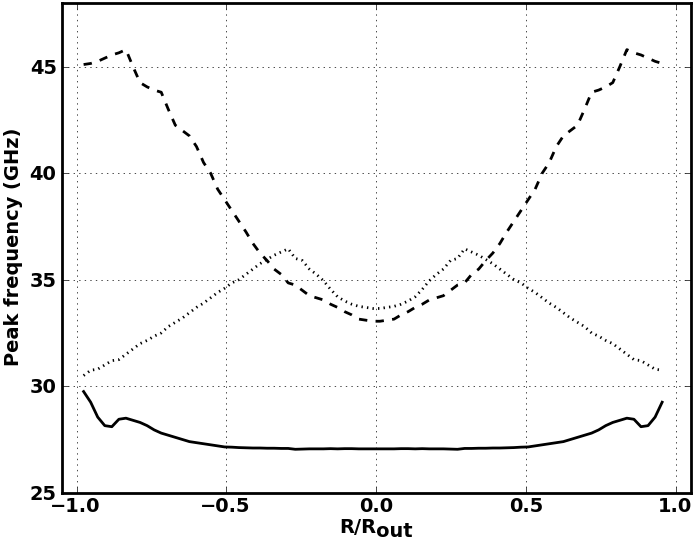}}
\caption{Peak frequency of the spinning dust emission for the molecular cloud with $n_0 = 10^3$ cm$^{-3}$ (dotted line), $10^4$ cm$^{-3}$ (dashed line), and $10^5$ cm$^{-3}$ (solid line).}
\label{peak_frequency_new} 
\end{figure}

We model a starless molecular cloud with an outer radius $R_{out} = 1$ pc and a central density $n_0 = 10^3$ H/cm$^3$. Fig. \ref{figure_1e3} shows the emerging dust emission for this cloud. Both the PAHs mid-IR emission and the spinning dust emission increase towards the centre of the cloud. However, the shapes of the profiles are quite dissimilar: the mid-IR profile is approximately twice as wide as the microwave profile. In this cloud, the intensity of the radiation field varies from 0.24 at the edge to 0.06 at the centre. With these low intensities, the dominant processes for the excitation and the damping of the spinning motion of the grains are the interactions with the gas. The spinning dust emission is consequently stronger at the centre of the cloud. On the other hand, because the PAH mid-IR emission is directly proportional to the intensity of the radiation field and to the column density, the grains emit more at the edge of the cloud, which produces a broader emission profile.

\subsection{Dense molecular globules: spinning dust as a tracer of grain growth from diffuse to dense medium}

\begin{figure}[!t]
\centerline{
\includegraphics[width=0.4\textwidth]{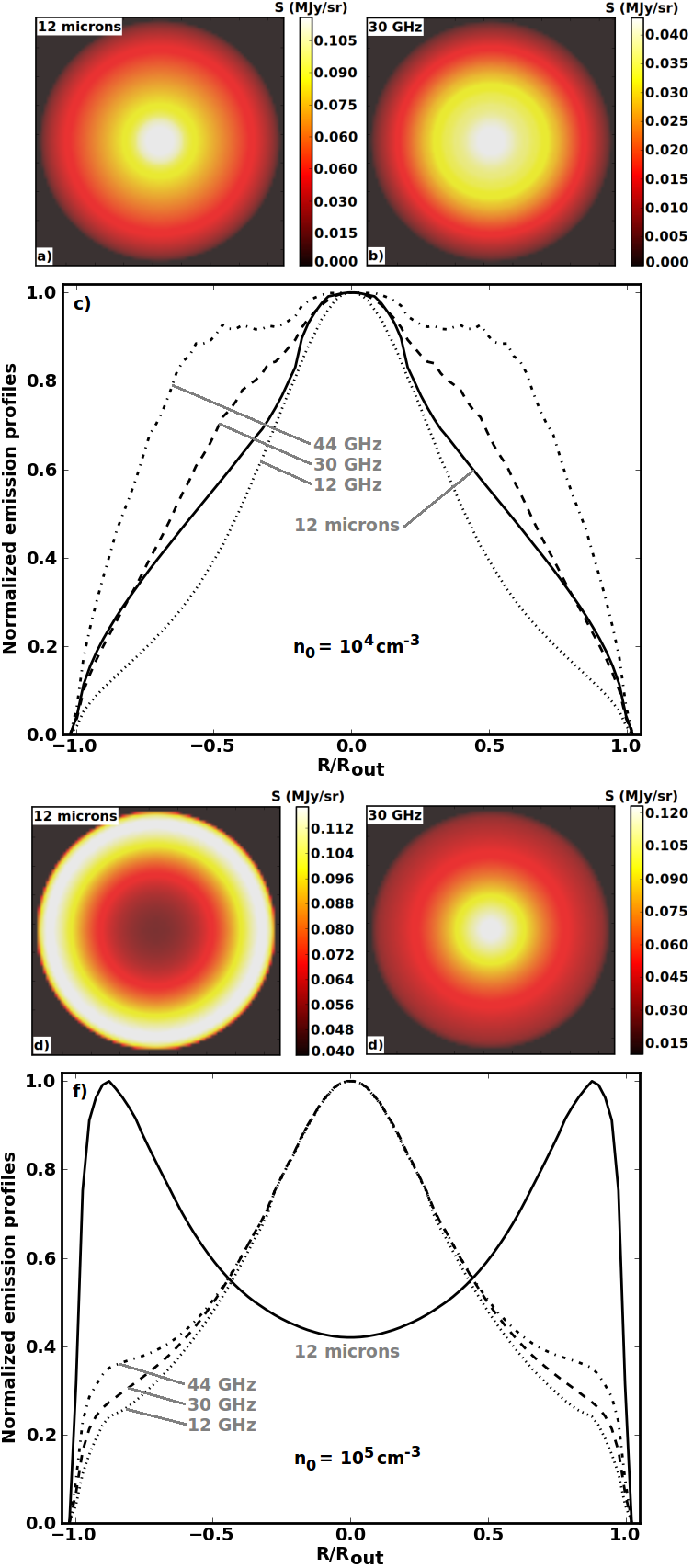}}
\caption{Surface brightness maps at 12 $\mu$m ($a$) and 30 GHz ($b$) and the emission profiles ($c$) at 12 $\mu$m (solid line), 12 GHz (dotted line), 30 GHz (dashed line), and 44 GHz (dot-dashed line) for the dense molecular globule with $n_0 = 10^4$ cm$^{-3}$. The three lower figures ($d$, $e$, $f$) show the same for $n_0 = 10^5$ cm$^{-3}$.}
\label{figure_transfert} 
\end{figure}

We modeled two dense molecular globules with an outer radius $R = 0.1$ pc and central densities $n_0 = 10^4$ and $10^5$ H/cm$^3$. The emerging dust emission at 12 $\mu$m, at the centre of two Planck-LFI filters (30, 44 GHz), and at 12 GHz are shown in Fig. \ref{figure_transfert}. As expected, the brightness maps in the mid-IR and in the microwave are dissimilar and even anti-correlated depending on the central density and on the frequency. The emission profiles in the microwave range have different widths depending on the frequency. This reflects the variation of the peak frequency of the spinning dust emission with the local physical conditions at a given radius (Fig. \ref{peak_frequency_new}). Apart from these general trends, the results are strikingly different depending on the central density of the cloud. We emphasize that these findings are quite different from what is expected and observed at degree scale in the diffuse interstellar medium where the mid-IR emission and the AME are correlated.

The cloud with $n_0 = 10^4$ H/cm$^3$ illustrates the complexity of the variations of the spinning dust emission with the environmental properties (Figs. \ref{figure_transfert}a-c). The peak frequency of the spinning dust emission varies from $\sim 45$ GHz at the edge to $\sim 33$ GHz at the centre of the cloud (Fig. \ref{peak_frequency_new}), while $0.1 \leqslant G_0 \leqslant 0.25$. These peak frequencies are higher than in the previous case with $n_0 = 10^3$ H/cm$^3$ (Section \ref{starless_molecular_cloud}), whereas the mid-IR emission is partly extinguished towards the centre. This explains why the profiles at 30 and 44 GHz are wider than the mid-IR profile. The width of the 12 GHz profile varies only little compared to the previous case because of its distance from the peak frequency.

For the densest cloud, with $n_0 = 10^5$ H/cm$^3$ (Figs. \ref{figure_transfert}d-f), the brightest region in the mid-IR is a ring at the edge of the cloud. The emission profile of these clouds can be explained in two ways. Either the PAHs are present throughout the cloud with a constant abundance but the radiation field is too extinguished at the centre for the grains to emit efficiently in the mid-IR. Some of the emitted IR photons also are absorbed by the dust, which additionally decreases the surface brightness at the centre. Alternatively the abundance of small grains decreases from the edge to the centre of the cloud. There is no way to distinguish these two scenarios with only data in the mid-IR. However, this ambiguity is lifted with a microwave emission map. Indeed, if PAHs are present at the centre of the cloud, the rotational excitation caused by the interactions with the gas particles is sufficient for them to emit in the microwave range. The peak frequencies vary from $\sim$ 30 GHz at the edge to $\sim$ 27 GHz at the centre, which coincides with the Planck-LFI bands and explains why the emission increases towards the cloud centre for the three frequencies. Because the dense clouds are transparent in the microwave range, spinning dust emission from the cloud centre can be observed. On the other hand, if small grains are absent from the centre, only a ring of mid-IR and spinning dust emission is observed at the edge of the cloud. As a result, spinning dust emission could be used to trace the evolution of small grains from the diffuse to the dense medium because the comparison between mid-IR and microwave data allow us to trace the variations of the PAHs abundance. In this context, the potential of comparing the AME to the mid-IR bands has recently been illustrated by \citet{Tibbs2011}.

\section{Conclusion}
\label{conclusions}

To understand the emerging spinning dust emission from interstellar clouds, we studied the influence of the gas ionisation state and of radiative transfer on the grain rotational motion with consistent models for the gas state and for the interstellar clouds. We found that the spinning dust spectrum is sensitive to the abundance of the major $\ion{H}{ii}$ and $\ion{C}{ii}$ ions and that these abundances must be described consistently in the modelling by including a treatment of the gas state. In addition, we showed that radiative transfer within interstellar clouds has surprising effects, e.g., that the mid-IR emission and the AME can be anticorrelated towards the cloud centre. From these findings, we argue that the AME from dense interstellar clouds provides new ways of grain growth from diffuse to dense medium (e.g., depletion of small grains by coagulation on larger ones), providing it is observed at relevant (a few arcminutes) angular scales using ground-based radio telescopes.

\acknowledgements{We thank our anonymous referee for useful comments and suggestions. N.Y. and M.J. acknowledge support from Academy of Finland project 127015.}

\bibliographystyle{aa} 
\bibliography{biblio}

\begin{thebibliography}{34}
\expandafter\ifx\csname natexlab\endcsname\relax\def\natexlab#1{#1}\fi

\bibitem[{{Ali-Ha{\"i}moud} {et~al.}(2009){Ali-Ha{\"i}moud}, {Hirata}, \&
  {Dickinson}}]{Ali2009}
{Ali-Ha{\"i}moud}, Y., {Hirata}, C.~M., \& {Dickinson}, C. 2009, \mnras, 395,
  1055

\bibitem[{{Arzoumanian} {et~al.}(2011){Arzoumanian}, {Andr{\'e}}, {Didelon},
  {K{\"o}nyves}, {Schneider}, {Men'shchikov}, {Sousbie}, {Zavagno}, {Bontemps},
  {di Francesco}, {Griffin}, {Hennemann}, {Hill}, {Kirk}, {Martin}, {Minier},
  {Molinari}, {Motte}, {Peretto}, {Pezzuto}, {Spinoglio}, {Ward-Thompson},
  {White}, \& {Wilson}}]{Arzoumanian2011}
{Arzoumanian}, D., {Andr{\'e}}, P., {Didelon}, P., {et~al.} 2011, \aap, 529,
  L6+

\bibitem[{{Battistelli} {et~al.}(2006){Battistelli}, {Rebolo},
  {Rubi{\~n}o-Mart{\'{\i}}n}, {Hildebrandt}, {Watson}, {Guti{\'e}rrez}, \&
  {Hoyland}}]{Battistelli2006}
{Battistelli}, E.~S., {Rebolo}, R., {Rubi{\~n}o-Mart{\'{\i}}n}, J.~A., {et~al.}
  2006, \apjl, 645, L141

\bibitem[{{Casassus} {et~al.}(2006){Casassus}, {Cabrera}, {F{\"o}rster},
  {Pearson}, {Readhead}, \& {Dickinson}}]{Casassus2006}
{Casassus}, S., {Cabrera}, G.~F., {F{\"o}rster}, F., {et~al.} 2006, \apj, 639,
  951

\bibitem[{{Casassus} {et~al.}(2008){Casassus}, {Dickinson}, {Cleary},
  {Paladini}, {Etxaluze}, {Lim}, {White}, {Burton}, {Indermuehle}, {Stahl}, \&
  {Roche}}]{Casassus2008}
{Casassus}, S., {Dickinson}, C., {Cleary}, K., {et~al.} 2008, \mnras, 391, 1075

\bibitem[{{Castellanos} {et~al.}(2011){Castellanos}, {Casassus}, {Dickinson},
  {Vidal}, {Paladini}, {Cleary}, {Davies}, {Davis}, {White}, \&
  {Taylor}}]{Castellanos2011}
{Castellanos}, P., {Casassus}, S., {Dickinson}, C., {et~al.} 2011, \mnras, 411,
  1137

\bibitem[{{Compi{\`e}gne} {et~al.}(2011){Compi{\`e}gne}, {Verstraete}, {Jones},
  {Bernard}, {Boulanger}, {Flagey}, {Le Bourlot}, {Paradis}, \&
  {Ysard}}]{Compiegne2011}
{Compi{\`e}gne}, M., {Verstraete}, L., {Jones}, A., {et~al.} 2011, \aap, 525,
  A103+

\bibitem[{{Dapp} \& {Basu}(2009)}]{Dapp2009}
{Dapp}, W.~B. \& {Basu}, S. 2009, \mnras, 395, 1092

\bibitem[{{Desert} {et~al.}(1990){Desert}, {Boulanger}, \&
  {Puget}}]{Desert1990}
{Desert}, F., {Boulanger}, F., \& {Puget}, J.~L. 1990, \aap, 237, 215

\bibitem[{{Draine} \& {Lazarian}(1998)}]{DL98}
{Draine}, B.~T. \& {Lazarian}, A. 1998, \apj, 508, 157

\bibitem[{{Ferland} {et~al.}(1998){Ferland}, {Korista}, {Verner}, {Ferguson},
  {Kingdon}, \& {Verner}}]{Ferland1998}
{Ferland}, G.~J., {Korista}, K.~T., {Verner}, D.~A., {et~al.} 1998, \pasp, 110,
  761

\bibitem[{{Goldsmith}(2001)}]{Goldsmith2001}
{Goldsmith}, P.~F. 2001, \apj, 557, 736

\bibitem[{{Hoang} {et~al.}(2010){Hoang}, {Draine}, \& {Lazarian}}]{Hoang2010}
{Hoang}, T., {Draine}, B.~T., \& {Lazarian}, A. 2010, \apj, 715, 1462

\bibitem[{{Juvela}(2005)}]{Juvela2005}
{Juvela}, M. 2005, \aap, 440, 531

\bibitem[{{Juvela} \& {Padoan}(2003)}]{Juvela2003}
{Juvela}, M. \& {Padoan}, P. 2003, \aap, 397, 201

\bibitem[{{Juvela} \& {Ysard}(2011)}]{Juvela2011}
{Juvela}, M. \& {Ysard}, N. 2011, [arXiv:astro-ph/1108.1345]

\bibitem[{{Kogut} {et~al.}(1996){Kogut}, {Banday}, {Bennett}, {Gorski},
  {Hinshaw}, \& {Reach}}]{Kogut1996}
{Kogut}, A., {Banday}, A.~J., {Bennett}, C.~L., {et~al.} 1996, \apj, 460, 1

\bibitem[{{Le Petit} {et~al.}(2006){Le Petit}, {Nehm{\'e}}, {Le Bourlot}, \&
  {Roueff}}]{LePetit2006}
{Le Petit}, F., {Nehm{\'e}}, C., {Le Bourlot}, J., \& {Roueff}, E. 2006, \apjs,
  164, 506

\bibitem[{{Leitch} {et~al.}(1997){Leitch}, {Readhead}, {Pearson}, \&
  {Myers}}]{Leitch1997}
{Leitch}, E.~M., {Readhead}, A.~C.~S., {Pearson}, T.~J., \& {Myers}, S.~T.
  1997, \apjl, 486, L23+

\bibitem[{{Liszt}(2003)}]{Liszt2003}
{Liszt}, H. 2003, \aap, 398, 621

\bibitem[{{L{\'o}pez-Caraballo} {et~al.}(2011){L{\'o}pez-Caraballo},
  {Rubi{\~n}o-Mart{\'{\i}}n}, {Rebolo}, \& {G{\'e}nova-Santos}}]{Lopez2011}
{L{\'o}pez-Caraballo}, C.~H., {Rubi{\~n}o-Mart{\'{\i}}n}, J.~A., {Rebolo}, R.,
  \& {G{\'e}nova-Santos}, R. 2011, \apj, 729, 25

\bibitem[{{Mathis} {et~al.}(1983){Mathis}, {Mezger}, \& {Panagia}}]{Mathis1983}
{Mathis}, J.~S., {Mezger}, P.~G., \& {Panagia}, N. 1983, \aap, 128, 212

\bibitem[{{McCall} {et~al.}(2003){McCall}, {Huneycutt}, {Saykally}, {Geballe},
  {Djuric}, {Dunn}, {Semaniak}, {Novotny}, {Al-Khalili}, {Ehlerding},
  {Hellberg}, {Kalhori}, {Neau}, {Thomas}, {{\"O}sterdahl}, \&
  {Larsson}}]{McCall2003}
{McCall}, B.~J., {Huneycutt}, A.~J., {Saykally}, R.~J., {et~al.} 2003, \nat,
  422, 500

\bibitem[{{Planck Collaboration} {et~al.}(2011{\natexlab{a}}){Planck
  Collaboration}, {Abergel}, {Ade}, {Aghanim}, {Arnaud}, {Ashdown}, {Aumont},
  {Baccigalupi}, {Balbi}, {Banday}, \& et~al.}]{PlanckMarshall2011}
{Planck Collaboration}, {Abergel}, A., {Ade}, P.~A.~R., {et~al.}
  2011{\natexlab{a}}, [arXiv:astro-ph/1101.2032]

\bibitem[{{Planck Collaboration} {et~al.}(2011{\natexlab{b}}){Planck
  Collaboration}, {Ade}, {Aghanim}, {Arnaud}, {Ashdown}, {Aumont},
  {Baccigalupi}, {Balbi}, {Banday}, {Barreiro}, \&
  et~al.}]{PlanckDickinson2011}
{Planck Collaboration}, {Ade}, P.~A.~R., {Aghanim}, N., {et~al.}
  2011{\natexlab{b}}, [arXiv:astro-ph/1101.2031]

\bibitem[{{R{\"o}llig} {et~al.}(2006){R{\"o}llig}, {Ossenkopf}, {Jeyakumar},
  {Stutzki}, \& {Sternberg}}]{Roellig2006}
{R{\"o}llig}, M., {Ossenkopf}, V., {Jeyakumar}, S., {Stutzki}, J., \&
  {Sternberg}, A. 2006, \aap, 451, 917

\bibitem[{{Shaw} {et~al.}(2006){Shaw}, {Ferland}, {Srianand}, \&
  {Abel}}]{Shaw2006}
{Shaw}, G., {Ferland}, G.~J., {Srianand}, R., \& {Abel}, N.~P. 2006, \apj, 639,
  941

\bibitem[{{Silsbee} {et~al.}(2011){Silsbee}, {Ali-Ha{\"i}moud}, \&
  {Hirata}}]{Silsbee2011}
{Silsbee}, K., {Ali-Ha{\"i}moud}, Y., \& {Hirata}, C.~M. 2011, \mnras, 411,
  2750

\bibitem[{{Tibbs} {et~al.}(2011){Tibbs}, {Flagey}, {Paladini}, {Compi{\'e}gne},
  {Shenoy}, {Carey}, {Noriega-Crespo}, {Dickinson}, {Ali-Ha{\"i}moud},
  {Casassus}, {Cleary}, {Davies}, {Davis}, {Hirata}, \& {Watson}}]{Tibbs2011}
{Tibbs}, C.~T., {Flagey}, N., {Paladini}, R., {et~al.} 2011,
  [arXiv:astro-ph/1108.2014]

\bibitem[{{Vidal} {et~al.}(2011){Vidal}, {Casassus}, {Dickinson}, {Witt},
  {Castellanos}, {Davies}, {Davis}, {Cabrera}, {Cleary}, {Allison}, {Bond},
  {Bronfman}, {Bustos}, {Jones}, {Paladini}, {Pearson}, {Readhead}, {Reeves},
  {Sievers}, \& {Taylor}}]{Vidal2011}
{Vidal}, M., {Casassus}, S., {Dickinson}, C., {et~al.} 2011, \mnras, 516

\bibitem[{{Williams} {et~al.}(1998){Williams}, {Bergin}, {Caselli}, {Myers}, \&
  {Plume}}]{Williams1998}
{Williams}, J.~P., {Bergin}, E.~A., {Caselli}, P., {Myers}, P.~C., \& {Plume},
  R. 1998, \apj, 503, 689

\bibitem[{{Wolfire} {et~al.}(2008){Wolfire}, {Tielens}, {Hollenbach}, \&
  {Kaufman}}]{Wolfire2008}
{Wolfire}, M.~G., {Tielens}, A.~G.~G.~M., {Hollenbach}, D., \& {Kaufman}, M.~J.
  2008, \apj, 680, 384

\bibitem[{{Ysard} {et~al.}(2010){Ysard}, {Miville-Desch{\^e}nes}, \&
  {Verstraete}}]{Ysard2010b}
{Ysard}, N., {Miville-Desch{\^e}nes}, M.~A., \& {Verstraete}, L. 2010, \aap,
  509, L1+

\bibitem[{{Ysard} \& {Verstraete}(2010)}]{Ysard2010a}
{Ysard}, N. \& {Verstraete}, L. 2010, \aap, 509, A12+

\end{thebibliography}

\end{document}